\documentclass[twocolumn,showpacs,preprintnumbers,superscriptaddress]{revtex4}
\usepackage{amsmath}
\usepackage{amssymb}
\usepackage{amsmath}
\usepackage{epsfig}
\usepackage{graphicx}
\usepackage{inputenc}
\inputencoding{latin1}
\usepackage{ulem}

\ifx\hypersetup\sadfkjashdfkxja\else
\hypersetup{pdftitle={A Proof of Factorization Theorem of Drell-Yan Process at Operator Level}}
\hypersetup{pdfauthor={Gao-Liang Zhou}}
\hypersetup{pdfsubject={}}
\hypersetup{pdfkeywords={}}
\hypersetup{plainpages=false}
\hypersetup{bookmarksnumbered=true}
\hypersetup{pdfstartview=FitH}
\hypersetup{pdfpagemode=UseNone}
\hypersetup{colorlinks=false}
\hypersetup{citebordercolor={.5 1 .5}}
\hypersetup{urlbordercolor={.5 1 1}}
\hypersetup{linkbordercolor={1 .7 .7}}
\fi

\newcommand{\ud}{\mathrm{d}}
\newcommand{\be}{\begin{equation}}
\newcommand{\ee}{\end{equation}}
\newcommand{\bea}{\begin{eqnarray}}
\newcommand{\eea}{\end{eqnarray}}

\newcommand{\Appendix}[1]%
    {%
     \section{#1}%
      }



\begin{document}

\title{A Proof of Factorization Theorem of Drell-Yan Process at Operator Level}
% \altaffiliation{}

\author{Gao-Liang Zhou
%\footnote{{\tt zhougl@itp.ac.cn}}
}
\affiliation{Key Laboratory of Frontiers in
Theoretical Physics, \\The Institute of Theoretical Physics, Chinese
Academy of Sciences, Beijing 100190, People's Republic of China}

%\email{zhougl@itp.ac.cn}
%\thanks{}


%\date{\today}

%%%%%%%%%%%%%%%%%%%%%%%%%%%%%%%%%%%%%%%%%%%%%%%%%%%%%%%%%%%%%%%%%%%%%%%%%%%%%%
\begin{abstract}
An alternative proof of factorization theorem for Drell-Yan process that works at operator level is given in this paper.  Interactions after the collision for such inclusive processes are proved to be canceled at operator level according to the unitarity of time evolution operator. After this cancellation, there are no longer leading pinch singular surface in Glauber region in the time evolution of electromagnetic currants. Decoupling of soft gluons from collinear jets is realized by defining new collinear fields that decouple from soft gluons. Cancelation of soft gluons is attribute to unitarity of time evolution operator and light-like Wilson lines of soft gluons.
\end{abstract}

\pacs{\it 12.38.-t, 12.39.St}
% 12.38.-t Quantum chromodynamics
% 12.39.St Factorization
\maketitle

%%%%%%%%%%%%%%%%%%%%%%%%%%%%%%%%%%%%%%%%%%%%%%%%%%%%%%%%%%%%%%%%%%%%%%%
\section{Introduction}

Factorization theorem of Drell-Yan process used to be a challenging question for a long time in past several decades \cite{BBL1981,M1982,LRS1982,CSS1984,B1985,CSS1985,CSS1988}. Being different from the cases in deep inelastic scattering process and annihilation process of $e^{+}e^{-}$, Glauber gluons can cause particular  difficulty in the proof of factorization theorem of Drell-Yan process.\cite{BBL1981,CSS1984,B1985,CSS1985,CSS1988}. It was proved that leading pinch singular surfaces(LPSS) caused by Glauber gluons will cancel out in covariant gauge after one sums over all possible cuts according to unitarity \cite{B1985,CSS1985,CSS1988}. Therefore, unitarity plays crucial role in the proof of factorization theorem of Drell-Yan process. The proofs appeared in \cite{B1985,CSS1985,CSS1988} work in diagram level and turn out to be complicated.

The hadronic tensor of Drell-Yan process reads:
\begin{equation}
\label{Hmunu}
H^{\mu\nu}=\int\ud^{4}x e^{-iq\cdot x}\big<p_{1}p_{2}|J^{\mu}(x)J^{\nu}(0)|p_{1}p_{2}\big>
\end{equation}
where $J^{\mu}(x)$ is the electromagnetic currant. It is connected with the differential cross section of Drell-Yan process by the formula:
\begin{equation}
\frac{\ud \sigma}{dq^{2}dy}
=\frac{4\pi\alpha^{2}Q_{q}^{2}}{3s}\int\frac{\ud^{2}\vec{q}_{\perp}}{(2\pi)^{4}}H^{\mu\nu}\frac{1}{q^{2}}
(\frac{q_{\mu}q_{\nu}}{q^{2}}-g_{\mu\nu})
\end{equation}
where $y=\frac{1}{2}\ln{q^{+}/q^{-}}$.

Momenta of initial hadrons become light-like in the high energy limit $Q\to\infty$, where $Q=(q^{2})^{1/2}$. hety are denoted as $p_{1}^{\mu}/Q=(p_{1}^{+},p_{1}^{-},\vec{p}_{1\perp})/Q\to(1,0,\vec{0})$ and $p_{2}^{\mu}/Q\to(0,1,\vec{0})$    in such limit where $p_{1}$ and $p_{2}$ represent momenta of initial hadrons. The degree of freedom  which contribute to leading pinch singularities, are:

(1)collinear particles, that is,  particles with momenta scale as $p^{2}/Q^{2}\to 0$ and $\ln|\frac{p^{0}+|\vec{p}|}{p^{0}-|\vec{p}|}|\to\infty$;

(2)hard particles, that is, particles with momenta scale as $p^{2}/Q^{2}\to a\neq 0$;

(3)soft particles, that is, particles with momenta scale as $p^{\mu}/Q\to 0$.

\begin{figure}
\begin{center}
\includegraphics[width=0.4\textwidth]{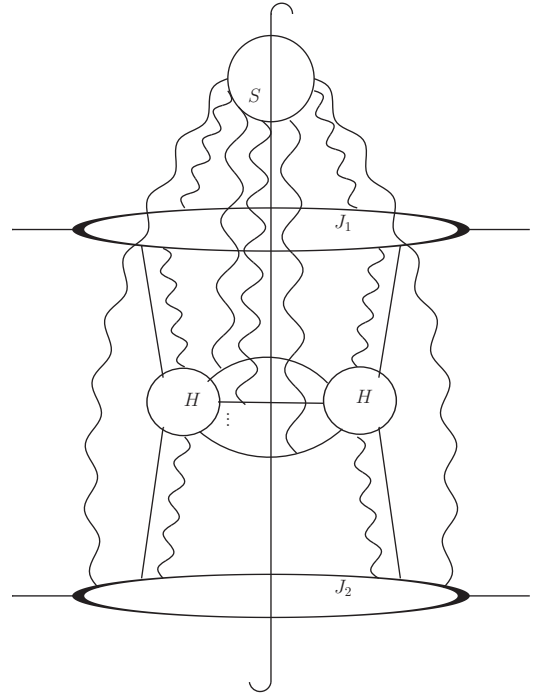}
\end{center}
\caption{The topology of diagrams that contain leading pinch singularities in Drell-Yan process}
\label{fig:pinch}
\end{figure}

Diagrams that contain leading pinch singularities have the same topology as in Fig.\ref{fig:pinch}\cite{B1985,CSS1985,CSS1988}.
There can be arbitrary number of soft gluons exchanged between collinear particles, which is denoted as $S$. Soft gluons do not connect to the hard vertex, which is denoted as $H$. There can also be arbitrary number of scalar-polarized gluons connected to the hard vertex for each jet in covariant gauge, which is proved to be appropriate for Drell-Yan process.\cite{B1985,CSS1985,CSS1988} There can be only one physical collinear particle(that is quark or transverse polarized gluons) connect to the hard vertex for each collinear jet.

To prove the factorization theorem one has to show that:

(1) Soft gluons exchanged between collinear particles do not violate factorization;

(2)Scalar polarized gluons connected to the hard vertex do not violate factorization.

Scalar polarized collinear gluons are absorbed into parton distribution function(PDF) according to Ward identity in \cite{B1985,CSS1985,CSS1988}. For the coupling between a soft gluon with momentum $q$ and a particle collinear-to-plus with momentum $p$ and mass $m$, people use the the eikonal line approximation (see for example \cite{CS1981B}):
\begin{equation}
A^{\mu}\simeq A^{-},\quad (p+q)^{2}-m^{2}\simeq 2p^{+}q^{-}+p^{2}-m^{2}
\end{equation}
to factorize the soft gluon from the collinear particle and finally prove the cancellation of soft gluons.
There is a special kind of soft gluons  can violate such approximation, which are often named as Glauber gluons in literatures.(\cite{BBL1981}) They are space like soft gluons with $\Lambda_{QCD}^{2}\lesssim|q^{2}|\ll Q^{2}$ and $|q|^{0}\ll E_{q}$. For example, one may consider the coupling between a particle collinear-to-plus and a soft gluon with momentum scale as $(q^{+},q^{-},\vec{q}_{\perp})/Q\sim (\lambda^{2},\lambda^{2},\vec{\lambda})$ where $\lambda\sim\Lambda_{QCD}/Q$, the eikonal line approximation does not work in such coupling.  For processes without initial hadrons moving in different direction, for example the annihilation of $e^{+}e^{-}$, there are not leading pinch singularities caused by Glauber gluons and one can deform the integral path to avoid such region. (\cite{CS1981B,CS1981S}). For processes with initial collinear hadrons moving in different directions, for example the Drell-Yan process, however, leading pinch singularities caused by Glauber gluons\cite{BBL1981} do exist and this can cause particular difficulty in the proof of factorization theorem for such processes.

To over come this difficulty, people\cite{B1985,CSS1985,CSS1988} sum over all possible final cuts in diagram level and show that there are no longer leading pinch singularities caused by Glauber gluons after this summation. One can then deform the integral path so that the eikonal line approximation works. Soft gluons factorize from collinear particles in this approximation and will cancel out.

In this paper, we notice that electromagnetic currants are color singlets and the summation over all possible final hadrons equals to summation over all possible final states in this process. Consequently,  unitarity can be implemented at operator level. With this insight, we present an alternative proof of factorization theorem of Drell-Yan process at operator level. This proof does not depend on explicit processes as it works at operator level and can be easily extend to some other processes. It also highlights theoretical aspects and physical pictures of factorization theorem more clearly compared to conventional diagram level approach.

We first show the cancellation of effects of interactions after the hard collision in (\ref{Hmunu}). This is the direct result of of time evolution operator of $QCD$. After such cancellation,
the $\delta$-function of energy conversation at the point $x_{i}$($x_{i}^{0}\leq x^{0}$) becomes $\int_{-\infty}^{x^{0}}\ud x_{i}^{0}e^{-ip_{i}^{0} x_{i}^{0}}\propto e^{-ip_{i}^{0} x^{0}}/(p_{i}^{0}+i\epsilon)$, where $p_{i}$ represent the total momenta that flow in to the vertex $x_{i}$, $x$ denote the point at which the hard photon is produced. If soft gluons(momenta $q$) couple to particles collinear-to-plus at the point $x_{i}$, then we can drop the plus momenta of soft gluons in the collinear internal lines. That is $q^{0}\propto q^{-}$  in the term $p_{i}^{0}+i\epsilon$. Thus singular points of $q^{-}$ produced by internal lines collinear-to-plus all locate in the lower half plane if we define $q$ as flow into the point $x_{i}$. This is our main strategy to prove the cancellation of LPSS caused by Glauber gluons. We will show this explicitly in this paper. The conclusion is that while dealing with coupling between a soft gluon with momenta $q$ and particles collinear-to-plus, one can always deform the integral path of $q^{-}$ to the upper half plane so that the eikonal line approximation works, where $q$ is defined as flow into the collinear particles. We work in the covariant gauge in such proof.

Coupling between soft gluons and collinear particles can then be absorbed into Wilson lines of scalar-polarized gluons. We also bring in effective hard vertex that describe the sub-processes. Effects of scalar-polarized collinear gluons are then absorbed into Wilson lines in the effective hard vertex according to gauge invariance. We then factorize and cancel effects of soft gluons at operator level in the hadronic tensor. Factorization theorem is then proved based on these facts.

The paper is organized as follows.
We prove the cancelation of effects of interactions after the collision in the hadronic tensor at operator level in Sec.\ref{cancellation}. In Sec.\ref{deformation}
We prove that one can deform the integral path so that the eikonal line approximation works in the hadronic tensor (\ref{Hmunu}).
In Sec.\ref{EFT}, we bring in the Wilson lines of scalar polarized gluons that absorb the effects of soft gluons. We also construct effective action that describe the electromagnetic scattering processes between different jets in this section.
In Sec.\ref{factorize}, we consider the hadronic tensor and finish the proof of  factorization theorem of Drell-Yan process.  Some discussions are given in Sec.\ref{conclusion}.

%%%%%%%%%%%%%%%%%%%%%%%%%%%%%%%%%%%%%%%%%%%%%%%%%%%%%%%%%%%%%%%%%%%%%%%
\section{Cancellation of effects of interactions after the collision}
\label{cancellation}
In this section we prove the cancellation of contributions of interactions after the collision in the hadronic tensor (\ref{Hmunu}). To make the proof clear, we start from the transition amplitude for an explicit process, which gives:
\begin{eqnarray}
&&_{out}\big<k_{1}\ldots k_{n}L|p_{1}p_{2}\big>_{in}
\nonumber\\
&=&\lim_{T\to\infty}\quad_{+}\big<k_{1}\ldots k_{n}L|e^{-iH(2T)}|p_{1}p_{2}\big>_{-}
\end{eqnarray}
where $|p_{i}\big>$ and $|k_{i}\big>$ denote hadrons, $L$ denote possible leptons or photons,  $|p_{1}p_{2}\big>_{-}$ and $_{+}\big<k_{1}\ldots k_{n}|$ are asymptotic states at times $t\to-\infty$ and $t\to\infty$ respectively. We are working in Schr$\ddot{o}$dinger picture in the right hand side of the equation.

There are not interactions between these particles at times $t\to\pm\infty$ according to adiabatic approximation.    We have:
\begin{equation}
|p_{1}p_{2}\big>_{-}=|p_{1}\big>_{-}|p_{2}\big>_{-}
\end{equation}
and
\begin{equation}
_{+}\big<k_{1}\ldots k_{n}L|=\quad_{+}\big<L|_{+}\big<k_{1}|\ldots \quad_{+}\big<k_{n}|
\end{equation}
As the consequence of confinement, such asymptotic states are separated hadrons(not quarks or gluons), leptons or photons. (The weak interaction is not concerned in this paper) We expand the asymptotic hadrons according to their parton contents:
\begin{equation}
|p_{i}\big>_{-}=\sum_{n,p_{i}^{j}}\quad _{-}\big<p_{i}^{1}\ldots p_{i}^{n}|p_{i}\big>_{-}|p_{i}^{1}\ldots p_{i}^{n}\big>_{-}
\end{equation}
and
\begin{equation}
|k_{i}\big>_{+}=\sum_{n,k_{i}^{j}}\quad _{+}\big<k_{i}^{1}\ldots k_{i}^{n}|k_{i}\big>_{+}|k_{i}^{1}\ldots k_{i}^{n}\big>_{+}
\end{equation}
where $|p_{i}^{j}\big>_{-}$ and $|k_{i}^{j}\big>_{+}$ represent parton states at $t\to-\infty$ and $t\to\infty$ respectively.
The transition amplitude can then be written as:
\begin{eqnarray}
&&\lim_{T\to\infty}\sum_{n_{i},m_{i}}\sum_{p_{i}^{j},k_{i}^{j}}
 \nonumber\\
&&
 _{-}\big<p_{1}^{1}\ldots p_{1}^{n_{1}}|p_{1}\big>_{-}\quad _{-}\big<p_{2}^{1}\ldots p_{2}^{n_{2}}|p_{2}\big>_{-}\nonumber\\
&& _{+}\big<k_{1}|k_{1}^{1}\ldots k_{1}^{m_{1}}\big>_{+}\ldots\quad _{+}\big<k_{n}|k_{n}^{1}\ldots k_{n}^{m_{n}}\big>_{+}
\nonumber\\
&&_{+}\big<k_{1}^{1}\ldots k_{1}^{m_{1}}\ldots k_{n}^{1}\ldots k_{n}^{m_{n}}L|
\nonumber\\
&&
e^{-iH(2T)}|p_{1}^{1}\ldots p_{1}^{m_{1}}p_{2}^{1}\ldots p_{2}^{m_{2}}\big>_{-}
\end{eqnarray}

For Drell-Yan process, we have:
\begin{equation}
H=H_{0}+(H_{int})_{QCD}+(H_{int})_{QED}
\end{equation}
where $(H_{int})_{QCD}$ and $(H_{int})_{QED}$ denotes interaction Hamiltonian of QCD and QED respectively. We keep the tree level result of QED and full result of QCD at this step and have:
\begin{eqnarray}
&&_{out}\big<k_{1}\ldots k_{n}L|p_{1}p_{2}\big>_{in}
\nonumber\\
&=&iQ_{q}e^{2}\lim_{T\to\infty}\int\ud^{4}x\int\frac{\ud^{4}q}{(2\pi)^{4}}e^{-iq\cdot x}
\nonumber\\
&&
_{+}\big<L|e^{-iH_{0}T}G^{\mu\nu}(q)j_{\nu}(q)e^{-iH_{0}T}|0\big>_{-}
\nonumber\\
&&_{+}\big<k_{1}\ldots k_{n}|e^{-iH_{QCD}(T-x^{0})}
\nonumber\\
&&
J_{\mu}(\vec{x})
e^{-iH_{QCD}(x^{0}+T)}
|p_{1}p_{2}\big>_{-}
\end{eqnarray}
where $Q_{q}$ is the electric-charge of the quark, $J^{\mu}$ and $j^{\mu}$ are the quark and lepton currants respectively, $G^{\mu\nu}(q)$ is the propagator of photons.  According to this transition amplitude, one can write the hadronic tensor (\ref{Hmunu}) as:
\begin{eqnarray}
H^{\mu\nu}(q)
&=&\lim_{T\to\infty}
 \int\ud^{4}x e^{-iq\cdot x} \quad_{-}\big<p_{1}p_{2}|
\nonumber\\
&&   e^{iH_{QCD}(x^{0}+T)}J^{\mu}(\vec{x})e^{iH_{QCD}(T-x^{0})}
\nonumber\\
&& e^{-iH_{QCD}T}J^{\nu}(\vec{0})e^{-iH_{QCD}T}|p_{1}p_{2}\big>_{-}
\end{eqnarray}

$H_{QCD}$ is Hermitian and we have:
\begin{eqnarray}
\label{hadronic tensor}
H^{\mu\nu}(q)
&=& \lim_{T\to\infty} \int_{-T}^{T}\ud x^{0}\int\ud^{3}\vec{x} e^{-iq\cdot x}
 \nonumber\\
&&
    _{-}\big<p_{1}p_{2}|
    e^{iH_{QCD}(x^{0}+T)}J^{\mu}(\vec{x})
\nonumber\\
&&
   e^{-iH_{QCD}x^{0}}J^{\nu}(\vec{0})e^{-iH_{QCD}T}
   |p_{1}p_{2}\big>_{-}
\nonumber\\
&=& \lim_{T\to\infty} \int_{-T}^{T}\ud x^{0}\int\ud^{3}\vec{x} e^{-iq\cdot x}
\quad _{-}\big<p_{1}p_{2}|
 \nonumber\\
&&
    e^{iH_{QCD}(x^{0}+T)}J^{\mu}(\vec{x})e^{-iH_{QCD}(x^{0}+T)}
\nonumber\\
&&
   e^{iH_{QCD}T}J^{\nu}(\vec{0})e^{-iH_{QCD}T}
   |p_{1}p_{2}\big>_{-}
\nonumber\\
&=&\int\ud^{4}x e^{-iq\cdot x}
\sum_{n_{i},m_{i}}\sum_{p_{i}^{j},k_{i}^{j},X}
\nonumber\\
&&_{-}\big<p_{1}^{1}\ldots p_{1}^{m_{1}}|p_{1}\big>_{-}\quad
   _{-}\big<p_{2}^{1}\ldots p_{2}^{m_{2}}|p_{2}\big>_{-}\nonumber\\
&& _{-}\big<p_{1}|k_{1}^{1}\ldots k_{1}^{n_{1}}\big>_{-}\quad
   _{-}\big<p_{2}|k_{2}^{1}\ldots k_{2}^{n_{2}}\big>_{-}
   \nonumber\\
&&
    _{0}\big<k_{1}^{1}\ldots k_{1}^{n_{1}}k_{2}^{1}\ldots k_{2}^{n_{2}}|
    U_{QCD}^{\dag}(x^{0},-\infty)
\nonumber\\
 &&   J^{\mu}(x)U_{QCD}(x^{0},-\infty)|X\big>_{0}
   \nonumber\\
&& _{0}\big<X|U_{QCD}^{\dag}(0,-\infty)J^{\nu}(0)
\nonumber\\
&& U_{QCD}(0,-\infty)|
  p_{1}^{1} \ldots p_{1}^{m_{1}}p_{2}^{1}\ldots p_{2}^{m_{2}}\big>_{0}
\end{eqnarray}
We have made use of unitaruty of the operator $e^{-iH_{QCD}(t_{1}-t_{2})}$ in the third line,  the last line works in the interaction picture. That is:
\begin{equation}
J^{\mu}(x)=e^{iH_{0}(x^{0}-t_{0})}J^{\mu}(\vec{x})e^{-iH_{0}(x^{0}-t_{0})}
\end{equation}
\begin{eqnarray}
&&U_{QCD}(t_{1},t_{2})
\nonumber\\
&=&e^{iH_{0}(t_{1}-t_{0})}e^{-iH_{QCD}(t_{1}-t_{2})}e^{-iH_{0}(t_{2}-t_{0})}
\nonumber\\
                    &=&T\exp\{-i\int_{t_{2}}^{t_{1}}\ud t(H_{I})_{QCD}(t)\}
\end{eqnarray}
with $t_{0}$ an arbitrary time. The parton sates $|X\big>_{0}$, $|p_{i}^{j}\big>_{0}$ and $|k_{i}^{j}\big>_{0}$ equal to corresponding  states in Schr$\ddot{o}$dinger picture at the time $t=t_{0}$.
$|p_{1}^{1} \ldots p_{2}^{m_{2}}\big>_{0}=|p_{1}^{1}\big>_{0}\ldots|p_{2}^{m_{2}}\big>_{0}$.  The time evolution operator $U_{QCD}(t_{1},t_{2})$ fulfills the identities:
\begin{equation}
U(t_{1},t_{2})U(t_{2},t_{3})=U(t_{1},t_{3})
\end{equation}
\begin{equation}
 U(t_{1},t_{2})U^{\dag}(t_{1},t_{2})=U^{\dag}(t_{1},t_{2})U(t_{1},t_{2})=1
\end{equation}

We see that there are not contributions of interactions after the collision in the second and third line of (\ref{hadronic tensor}).

%%%%%%%%%%%%%%%%%%%%%%%%%%%%%%%%%%%%%%%%%%%%%%%%%%%%%%%%%%%%%%%%%
\section{Deformation of integral path}
\label{deformation}

We now prove that we can deform the integral path so that the eikonal line approximation works in (\ref{Hmunu}).  Our conclusion is that while dealing with the coupling between soft gluons(momenta $q$) and particles collinear-to-plus, one can always deform the integral path of $q^{-}$ to the upper half plane so that the eikonal line approximation works, where $q$ is defined as flow into the collinear particles. We work in covariant gauge in this section.

We start from  the evolution:
\begin{eqnarray}
&&U_{QCD}^{\dag}(x^{0},-\infty)J^{\mu}(x)U_{QCD}(x^{0},-\infty)
\nonumber\\
&=&U_{QCD}^{\dag}(x^{0},-\infty)\bar{\psi}(x)U_{QCD}(x^{0},-\infty)\gamma^{\mu}\nonumber\\
&&U_{QCD}^{\dag}(x^{0},-\infty)\psi(x)U_{QCD}(x^{0},-\infty)
\end{eqnarray}
with $|x^{0}|\lesssim 1/Q$. The evolution of the currant is factorized into product of evolution of fermion fields. We first look at the evolution of the field $\psi{x}$. According to standard perturbation calculations,
we have:
\begin{eqnarray}
\label{Glauber}
&&U_{QCD}^{\dag}(x^{0},-\infty)\psi(x)U_{QCD}(x^{0},-\infty)\nonumber\\
&=&\psi(x)+\sum_{n}(-i)^{n}\int_{t_{2}}^{x^{0}}\ud t_{1}\ldots\int_{-\infty}^{x^{0}}\ud t_{n}
\nonumber\\
&&[\ldots[\psi(x),H_{I_{QCD}}(t_{1})],\ldots,H_{I_{QCD}}(t_{n})]
\nonumber\\
&=&\psi(x)+\sum_{n}(-g)(-i)^{n}
\nonumber\\
&&\int_{t_{2}}^{x^{0}}\ud t_{1}\ldots\int_{-\infty}^{x^{0}}\ud t_{n}
\int\ud^{3}\vec{x_{1}}G_{R}(x,x_{1})
\nonumber\\
&&
[\ldots[\not\!A(x_{1})\psi(x_{1}),H_{I_{QCD}}(t_{2})],\ldots,
\nonumber\\
&&
H_{I_{QCD}}(t_{n})]
\nonumber\\
\end{eqnarray}
where
\begin{eqnarray}
&&G_{R}^{ij}(x,x_{1})
\nonumber\\
&=&\theta(x^{0}-x_{1}^{0})[\psi(x)^{r},\bar{\psi}^{s}(x_{1})]
\nonumber\\
&=&\int\frac{\ud^{4}k}{(2\pi)^{4}}
    \frac{i(\not\!k\pm m)^{ij}e^{-ik\cdot(x-x_{1})}}{(k^{0}-E_{k}+i\epsilon)(k^{0}+E_{k}+i\epsilon)}
\end{eqnarray}
\begin{eqnarray}
&&G_{R}^{\mu\nu}(x,x_{1})
\nonumber\\
&=&\theta(x^{0}-x_{1}^{0})[A^{\mu}(x),A^{\nu}(x_{1})]
\nonumber\\
&=&
\int\frac{\ud^{4}k}{(2\pi)^{4}}
    \frac{-i e^{-ik\cdot(x-x_{1})}}{(k^{0}-E_{k}+i\epsilon)(k^{0}+E_{k}+i\epsilon)}
\nonumber\\
&&(g^{\mu\nu}-\frac{(1-\xi)k^{\mu}k^{\nu}}{(k^{0}-E_{k}+i\epsilon)(k^{0}+E_{k}+i\epsilon)})
\end{eqnarray}
One should pay attention to the sign of the $i\epsilon$ in these propagators caused by the step function $\theta(x^{0}-x_{1}^{0})$.  We repeat the calculation and get the result:
\begin{eqnarray}
\label{xspace}
&&U_{QCD}^{\dag}(x^{0},-\infty)\psi(x)U_{QCD}(x^{0},-\infty)
\nonumber\\
&=&\psi(x)+\sum_{n}\sum_{1\leq i<j\leq n}\sum_{G_{R},V_{n}^{i}} (-i)^{n}
\nonumber\\
&&
\int_{-\infty}^{x^{0}}\ud t_{1}\ldots\int_{-\infty}^{t_{n-1}}\ud t_{n}
\int\ud^{3}\vec{x}_{1}\ldots\int\ud^{3}\vec{x}_{n}
\nonumber\\
&&G_{R}(x,x_{1})\ldots G_{R}(x_{i},x_{j})\ldots G_{R}(x_{n-1},x_{n})
\nonumber\\
&&
V_{n}^{1}(\psi,\bar{\psi},A^{\mu})(x_{1})\ldots V_{n}^{n}(\psi,\bar{\psi},A^{\mu})(x_{n})
\end{eqnarray}
where $V_{n}^{i}$ are function of fermion fields and gluon fields that are not contracted with other fields. For example, $V_{1}^{1}=-g\not\!{A}(x_{1})\psi(x_{1})$. $G_{R}(x_{i},x_{j})$ equals to $G_{R}^{rs}(x_{i},x_{j})$ for fermions and $G_{R}^{\mu\nu}(x_{i},x_{j})$ for gluons, $x_{0}=x$. We do not contract fields in $V_{n}^{i}$ at this step, thus there is no more than one internal line connect $x_{i}$ and $x_{j}$ direct for each pair of $x_{i}$ and $x_{j}$.  One should sum over all possible combination of $V_{n}^{i}$ and $G_{R}$, this is suggested by the third summation. If there are two points $x_{i}$ and $x_{j}$($i<j$) not connected by the chains $x_{i}\to x_{i_{1}}\ldots\to x_{j}$  then the integral function in above formula is invariant under the permutation $x_{i}\leftrightarrow x_{j}$. This is because that we sum over all possible combination of $V_{n}^{i}$ and $G_{R}$, internal lines or fields that connect to $x_{i}$ can also connect to $x_{j}$. Thus the order of $x_{i}^{0}$ and $x_{j}^{0}$ do not affect result in such case. We also notice that the propagators $G_{R}(x_{l},x_{m})=0$ if $x_{l}^{0}<x_{m}^{0}$, thus we can drop the constraints that $x_{m}^{0}<x_{l}^{0}$ and have:
\begin{eqnarray}
&&\psi(x)+\sum_{n}\sum_{1\leq i<j\leq n}\sum_{G_{R},V_{n}^{i}}C(G_{R},V_{n}^{i}) (-i)^{n}
\nonumber\\
&&
\int_{-\infty}^{x^{0}}\ud t_{1}\ldots\int_{-\infty}^{x^{0}}\ud t_{n}
\int\ud^{3}\vec{x}_{1}\ldots\int\ud^{3}\vec{x}_{n}
\nonumber\\
&&G_{R}(x,x_{1})\ldots G_{R}(x_{i},x_{j})\ldots G_{R}(x_{n-1},x_{n})
\nonumber\\
&&
V_{n}^{1}(\psi,\bar{\psi},A^{\mu})(x_{1})\ldots V_{n}^{n}(\psi,\bar{\psi},A^{\mu})(x_{n})
\end{eqnarray}
where $C(G_{R},V_{n}^{i})$ denote possible symmetrization factors caused by the permutation $x_{i}\leftrightarrow x_{j}$, which are not connected by the chains $x_{i}\to x_{i_{1}}\ldots\to x_{j}$ or $x_{j}\to x_{j_{1}}\ldots\to x_{i}$.

For the evolution of $\bar{\psi}$ we have similar results. To get the matrix-element of the evolution of the currants we only need to contract fields in $V_{n}^{i}$ terms with other fields in such terms or contract them with initial or final states.

We write the above formula in momenta space:
\begin{eqnarray}
\label{Mspace}
&&\psi(x)+\sum_{n}\sum_{1\leq i<j\leq n}\sum_{G_{R},V_{n}^{i}}(-i)^{n}
\nonumber\\
&&
(\prod_{i=1}^{n}\int\frac{\ud^{4}q_{i}}{(2\pi)^{4}})
(\prod_{1\leq i\leq n-1}^{i<j\leq n}\int\frac{\ud^{4}k_{ij}}{(2\pi)^{4}})
\nonumber\\
&&e^{-i\sum_{i=1}^{n}q\cdot x}(\prod_{1\leq i\leq n-1}^{i<j\leq n}G_{R}(k_{ij}))
(\prod_{i=1}^{n}V_{n}^{i}(q_{i}))
\nonumber\\
&&
(\prod_{i=1}^{n}(2\pi)^{3}\delta^{(3)}(\vec{q}_{i}+\sum_{i<j\leq n}\vec{k}_{ij}-\sum_{1\leq j<i}\vec{k}_{ji}))
\nonumber\\
&&(\prod_{i=1}^{n}\frac{1}{-i(\sum_{j=i}^{n}q_{j}^{0}
-\sum_{i\leq j\leq n}^{1\leq l< i}k_{lj}^{0}+i\epsilon)})
\end{eqnarray}
or
\begin{eqnarray}
\label{Mspace2}
&&
\psi(x)+\sum_{n}\sum_{1\leq i<j\leq n}\sum_{G_{R},V_{n}^{i}}C(G_{R},V_{n}^{i}) (-i)^{n}
\nonumber\\
&&(\prod_{i=1}^{n}\int\frac{\ud^{4}q_{i}}{(2\pi)^{4}})
(\prod_{1\leq i\leq n-1}^{i<j\leq n}\int\frac{\ud^{4}k_{ij}}{(2\pi)^{4}})
e^{-i\sum_{i=1}^{n}q\cdot x}
\nonumber\\
&&
(\prod_{i=1}^{n}V_{n}^{i}(q_{i}))(\prod_{1\leq i\leq n-1}^{i<j\leq n}G_{R}(k_{ij}))
\nonumber\\
&&
(\prod_{i=1}^{n}
(2\pi)^{3}\delta^{(3)}(\vec{q}_{i}+\sum_{i<j\leq n}\vec{k}_{ij}-\sum_{1\leq j<i}\vec{k}_{ji}))
\nonumber\\
&&
(\prod_{i=1}^{n}(\frac{i}{q_{i}^{0}+\sum_{i<j\leq n}k_{ij}^{0}-\sum_{1\leq j<i}k_{ji}^{0}+i\epsilon}))
\end{eqnarray}

We now consider the coupling between Glauber gluons and other modes. In the coupling between Glauber gluon(momenta $q_{G}$) and soft(including Glauber) particles(momenta $q_{s}$),  there are at least one Glauber gluon emitted or absorbed during the time range of propagation of a soft particle if $E_{\vec{q}_{s}}\gtrsim E_{\vec{q}_{G}}$, this is suppressed as power of $q_{G}^{0}/E_{\vec{q}_{s}}$. If $E_{\vec{q}_{s}}\ll E_{\vec{q}_{G}}$ then there are at least one (ultra-)soft particle emitted or absorbed during the time range of propagation of a Glauber gluon, this is suppressed as power of $\min\{1,|q_{G}^{0}/E_{\vec{q}_{s}}|\}E_{\vec{q}_{s}}/E_{\vec{q}_{G}}$. Thus one can take the eikonal line approximation in the coupling between Glauber gluons and collinear particles with corrections of order $E_{\vec{q}_{G}}/p^{0}$ if the other end of Glauber gluons are not collinear particles, where $p$ represent momenta of collinear particles. We only need to consider the case that both ends of the Glauber gluons are collinear particle.
In  the coupling between soft (including Glauber) gluons and particles collinear-to-plus, the soft gluons should be polarized in the minus direction at leading order. While working in covariant gauge, the propagator
\begin{equation}
\frac{-i}{(q_{i}^{s})^{2}}(g^{\mu\nu}-(1-\xi)\frac{q_{i}^{s\mu}q_{i}^{s\nu}}{(q_{i}^{s})^{2}})
\end{equation}
do not contain singular points locate in Glauber region.  We notice that $|q_{i}^{s-}q_{i}^{s\mu}|\ll|\vec{q}_{i\perp}^{s}|^{2}$ for arbitrary $\mu$ if $q_{i}^{s}$ locate in Glauber region. Thus, to give leading order contribution the other end of the propagators of Glauber gluons should be polarized in the plus direction. As the result, the other end should be collinear particles with large minus momenta.

If there are not collinear internal lines at a point $x_{i}$, that is, all collinear fields in $V_{n}^{i}$ are contracted with initial or final states and all $k_{ij}$ and $k_{li}$($l<i<j$) are not collinear, then we can neglect the small momenta components of the initial and final particles in the wave function($u(p)$ or $v(p)$) or the polarization vectors.  We will assume that there are at least one collinear internal line at the vertex $x_{i}$ at which soft particles couple to collinear particles.

We consider the case that soft gluons couple to particles collinear-to-plus at the point $x_{i}$ and have:
\begin{eqnarray}
\label{momenta coversation}
&&\frac{\delta(q_{i}^{3}+\sum_{i<i^{\prime}\leq n}k_{ii^{\prime}}^{3}-\sum_{1\leq i^{\prime}<i}k_{i^{\prime}i}^{3})}{q_{i}^{0}+\sum_{i<j^{\prime}\leq n}k_{ij^{\prime}}^{0}-\sum_{1\leq j^{\prime}<i}k_{j^{\prime}i}^{0}+i\epsilon}
\nonumber\\
&=&
\frac{1}{\sqrt{2}}\frac{\delta(q_{i}^{3}+\sum_{i<i^{\prime}\leq n}k_{ii^{\prime}}^{3}-\sum_{1\leq i^{\prime}<i}k_{i^{\prime}i}^{3})}{q_{i}^{-}+\sum_{i<j^{\prime}\leq n}k_{ij^{\prime}}^{-}-\sum_{1\leq j^{\prime}<i}k_{j^{\prime}i}^{-}+i\epsilon}
\end{eqnarray}
We then make the approximation:
\begin{eqnarray}
\label{colliniear approximation}
&&\delta(q_{i}^{3}+\sum_{i<i^{\prime}\leq n}k_{ii^{\prime}}^{3}-\sum_{1\leq i^{\prime}<i}k_{i^{\prime}i}^{3})
\nonumber\\
&=&
\sqrt{2}\delta((q_{i}^{+}+q_{i}^{-})+\sum_{i<i^{\prime}\leq n}(k_{ii^{\prime}}^{+}
+k_{ii^{\prime}}^{-})
\nonumber\\
&&
-\sum_{1\leq i^{\prime}<i}(k_{i^{\prime}i}^{+}+k_{i^{\prime}i}^{-}))
\nonumber\\
&\simeq& \sqrt{2}\delta(\widetilde{q}_{i}^{+}+\sum_{i<i^{\prime}\leq n}\widetilde{k}_{ii^{\prime}}^{+}-\sum_{1\leq i^{\prime}<i}\widetilde{k}_{i^{\prime}i}^{+})
\end{eqnarray}
 where $\widetilde{p}_{i}=p_{i}$ for collinear particles, $\widetilde{p}_{i}=(0,p_{i}^{-},(\vec{p}_{i})_{\perp})$ for soft particles. We can make such approximation as there are collinear internal lines. For internal lines collinear-to-plus only the large components of $l^{+}$ is important.

If there are soft gluons in the $V_{n}^{i}$, the momenta of which we denote as $q_{i}^{s}$, then singular point of $q_{i}^{s-}$ that locate in the upper half plane can only be produced by the $V_{n}^{i}$ term in (\ref{Mspace2}) after (\ref{momenta coversation}) and (\ref{colliniear approximation}). In addition, (\ref{Mspace2}) can rely on $q_{i}^{s+}$ only through the $V_{n}^{i}$ term.  We also see that singular points of $k_{ij}^{-}$($j>i$) in Glauber region that locate in the upper half plane can only be produced by the other end of $k_{ij}$, while that of $k_{i^{\prime}i}^{-}$($i^{\prime}i$) that locate in the lower half plane can only be produced by the other end of $k_{i^{\prime}i}$.

If the other end of soft gluons connect to particles collinear to $n^{\mu}$ with $n^{3}=\cos(\theta)$, then singular point of $q_{s}$($q_{s}=q_{i}^{s}$, $k_{ij}$ or $k_{i^{\prime}i}$) produced by collinear internal lines at that end are those $n\cdot q_{s}\sim |(\vec{q}_{s})_{n\perp}^{2}|/Q$, where $\vec{n}_{\perp}$ denote the vector that fulfill the condition $\vec{n}_{\perp}\cdot\vec{n}=0$. We can then deform the integral path of $q_{i}^{s-}$ and $k_{ij}^{-}$ to upper half plane and that of $k_{i^{\prime}i}^{-}$ to lower half plane with radius of order:
 \begin{equation}
\min\{\frac{|(\vec{q}_{s})_{\perp}||\sin(\theta)|}{(1+\cos(\theta))},|\vec{q}_{s}|\}
\end{equation}. After this deformation, we can drop the components $(\vec{q}_{s})_{\perp}$ in collinear internal lines at the point $x_{i}$ with corrections suppressed as power of
\begin{equation}
\min\{\max\{\frac{|(\vec{q}_{s})_{\perp}|(1+\cos(\theta))}{Q|\sin(\theta)|},\frac{|\vec{q}_{s}|}{Q}\},
\frac{|\vec{q}_{s}|^{2}}{Q{q}_{s}^{-}}\}
 \end{equation}
We also notice that  such coupling is itself of order $\max\{(1-\cos(\theta)),(1-\xi)|{q}_{s}^{-}|/|\vec{q}_{s}|\}$ as $n^{-}=1-\cos(\theta)$
, where $\xi$ the is gauge parameter. The corrections to the approximation are no greater than:
\begin{equation}
\max\{\frac{|(\vec{q}_{s})_{\perp}|(1+\cos(\theta))}{Q|\sin(\theta)|}\ast (1-\cos(\theta)),
\frac{|\vec{q}_{s}|}{Q}\}\lesssim \frac{|\vec{q}_{s}|}{Q}.
 \end{equation}
where we have assumed that $1-\xi$ is not too large, this is the case in the Feynman gauge and Landau gauge.
We notice that $q_{i}$ and $k_{ij}$ are defined as flow in to the point $x_{i}$, $k_{i^{\prime}i}$ is defined as flow out of the point $x_{i}$, such deformation is in accordance with our claim at the beginning of this section. After such deformation, we can drop the the components $(\vec{q}_{s})_{\perp}$ in collinear internal lines at the pint $x_{i}$ with corrections of order $|\vec{q}_{s}|/Q$. That is, the eikonal line approximation does work after this deformation.

Our proof do not rely on the explicit form of $V_{n}^{i}(x_{i})$ and can be extended to such evolution  of other operators that are local in time. One can easily see that, our proof can also be extend to the case that  interactions before the collision are cancelled. In that case, one should define that $x_{i}^{0}<x_{j}^{0}$ for $i<j$. One should change the $i\epsilon$ term to $-i\epsilon$  in (\ref{Mspace2}).   Then according to similar proof, one can see that while dealing with the coupling between Glauber gluons(momenta $q$) and particles collinear-to-plus, one can always deform the integral path of $q^{-}$ to the lower half plane so that the eikonal line approximation works.

%%%%%%%%%%%%%%%%%%%%%%%%%%%%%%%%%%%%%%%%%%%%%%%%%%%%%%%%%%%%%%%%%
\section{Wilson lines of collinear and soft gluons}
\label{EFT}

We have proved the cancellation of leading pinch singularities caused by Glauber gluons and deformed the integral path to avoid the Glauber region. We can now use the eikonal line approximation in the coupling between soft gluons and collinear particles. We bring in Wilson lines of scalar-polarized gluons to absorb the effects of soft gluons  in this section. We also bring in effective action equipped with Wilson lines of scalar-polarized collinear gluons to describe the hard process in this section.

It is convenient to first determine the power counting for different modes. For a massless scalar field $\phi(p)$ with momenta scale as $p^{\mu}\sim Q\lambda$, we have:
\begin{equation}
[\phi(p),\phi(p^{\prime})]=\frac{i}{p^{2}+i\epsilon}\delta^{(4)}(p-p^{\prime})
\end{equation}
Power counting for $\phi(x)$ reads:
\begin{equation}
\phi(p) \sim (Q\lambda)^{-3}
\end{equation}
Power counting for other modes  can be determined according to similar method. We give the result in covariant gauge in Table \ref{Power couting},
\begin{table*}
\caption{Power counting for various modes in covariant gauge}
\begin{tabular}{ccc}
\hline
Modes & Momenta$(\bar{n}\cdot p, n\cdot p, \vec{p}_{n\perp})$ & Power counting\\
\hline
$\psi(p)$ & $Q(1,\lambda_{n},\vec{\lambda}_{n\perp})$ & $Q^{-5/2}\lambda_{min}^{-1/2}\lambda_{max}^{-1}$\\
\hline
$A^{\mu}(p)$ & $ Q(1,\lambda_{n},\vec{\lambda}_{n\perp})$
& $Q^{-3}\lambda_{min}^{-1/2}\lambda_{max}^{-1}
(\lambda_{max}^{-1/2}, 1,\lambda_{max}^{1/2})$\\
\hline
$\psi(p)$ & $p^{\mu}\sim Q\lambda_{s}$ & $(Q\lambda_{s})^{-5/2}$\\
\hline
$A^{\mu}(p)$ & $p^{\mu}\sim Q\lambda_{s}$ & $(Q\lambda_{s})^{-3}$\\
\hline
\end{tabular}
\label{Power couting}
\end{table*}
where $n^{\mu}=\frac{1}{\sqrt{2}}(1,\vec{n})$ is a light-like direction with $|\vec{n}|^{2}=1$, $\vec{p}_{n\perp}\cdot \vec{n}=0$, $\bar{n}^{\mu}=\sqrt{2}(1,\vec{0})-n^{\mu}$, $\lambda_{n}$, $\lambda_{n\perp}$ and $\lambda_{s}$ are parameters much smaller than 1, $\lambda_{min}=\min\{\lambda_{n},\lambda_{n\perp}^{2}\}$, $\lambda_{max}=\max\{\lambda_{n},\lambda_{n\perp}^{2}\}$. There are ambiguities in power counting of collinear gluons. If the parameter $1-\xi$ is of order $1$, one can easily get the result in Table \ref{Power couting}. If $1-\xi\ll 1$ then we have $g^{nn}=g^{\bar{n}\bar{n}}=0$ and $g^{n\bar{n}}=g^{\bar{n}n}=1$. Thus, power counting for $\bar{n}\cdot A$ and $n\cdot A$ is not definite in this case. We have determined the power counting by requiring that power counting for coupling between collinear fermions and collinear gluons is $\int\ud^{4}x\bar{\psi} \not\!A\psi\sim 1$.  We do not show the power counting for Glauber gluons as one can deform the integral path to avoid Glauber region.  The summation over $n^{\mu}$,  $\sum_{n^{\mu}}$  is of order $\theta^{2}/\theta_{n}^{2}$,  where $\theta_{n}$ is the angle resolution of the directions $n^{\mu}$, $\theta$ is the typical scattering angle of the process. For Drell-Yan process, we have $\theta\sim 1$. We also notice that for the contraction of two fields with momenta in the collinear region to be non-zero, the momenta of these two fields should collinear to the same direction. Power counting for $\sum_{n^{\mu}}$ then reads $\sum_{n^{\mu}}\sim\theta/\theta_{n}$. We do not absorb this into power counting of fields, as there can be at most one such summation for each collinear jet.

We consider a scattering process between collinear particles with scattering angle of order $\theta$. $\theta$ is constrained to be $M/Q\lesssim \theta\lesssim 1$, where $M$ is the typical mass scale of hadrons that appear in the process. For internal lines of this process, we have $|k^{2}|\sim Q^{2}\theta^{2}$. For collinear external lines we have $|k_{n}|^{2}\sim Q^{2}\theta^{2}\lambda_{n}^{2}$ with $\lambda_{n}\lesssim 1$. There can be soft gluons participate in the process, for which we have $|k^{2}|\sim  Q^{2}\theta^{2} \lambda_{s}^{2}$ with $\lambda_{s}\lesssim 1$. We assume that soft gluons connect to the internal lines as process with soft gluons connect to the external lines should be treated as two subprocesses.  There are not super-renormalizable interactions in QCD and the power of $\theta$ reads $\theta^{D}$ for this process with $D\geq 0$. The power of $\lambda_{s}$ is produced by the term $\int\ud^{4} q A_{s}(q)$, which reads $\lambda_{s}^{S}$ with $S$ the number of soft gluons. The power of $\lambda_{n}$ is produced by the term $\int\ud^{4}p\psi(p)$ and $\int\ud^{4}pA(p)$ and the summation $\sum_{n^{\mu}}$, which reads $\lambda_{n}^{d_{n}-1}=\lambda_{n}^{\sum_{i}d_{n}^{1}-1}$, where $d_{n}^{i}=1$ for collinear fermions, $d_{n}^{i}=0,1,2$ for $\bar{n}\cdot A_{n}$, $A_{n\perp}$ and $n\cdot A_{n}$ respectively. Power counting for the process reads $\theta^{D}\lambda_{s}^{S}\prod_{n^{\mu}}\lambda_{n}^{d_{n}-1}$. If $d_{n}\geq 1$ then there not super leading powers produced by particle collinear to $n^{\mu}$. If all particles collinear to $n^{\mu}$ are scalar polarized gluons then according to Ward identity, there is an additional power of $\theta^{2}\lambda_{n}^{2}$ and the super-leading powers cancel out.

Coupling between particles collinear to $n^{\mu}$ and soft fermions are of order $(|n\cdot  p||q^{\mu}|)^{1/2}/(\bar{n}\cdot p)$ according to the power counting displayed in Table \ref{Power couting}, where $p$ and $q$ denotes momenta of collinear particles and soft fermions. We neglect such coupling in this paper.

We now consider the classical configuration of parton fields in the region $S(x,1/\Lambda_{QCD})-H(x,1/Q)$. Different jets separate from each other in this region although they exchange soft gluons in this region. We denote the region which jet collinear to $n^{\mu}$ locate in as $y_{n}$. We also denote the region in which there are nt collinear jets as $y_{s}$.
($y_{n}\subset S(x,1/\Lambda_{QCD})-H(x,1/Q)$, $y_{s}\subset S(x,1/\Lambda_{QCD})-H(x,1/Q)$). In the region $y_{s}$, we can define the effective fields:
\begin{equation}
\psi_{s}(y_{s})=\psi(y_{s})
,\quad
A_{s}^{\mu}(y_{s})=A^{\mu}(y_{s})
\end{equation}
We extend the definition of soft fields into the region $y_{n}$ according to the manner:
\begin{equation}
(D_{s\mu}G_{s}^{\mu\nu})^{a}(y_{n})=g\bar{\psi}_{s}\gamma^{\mu}t^{a}\psi_{s}(y_{n})
\end{equation}
\begin{equation}
 A_{s}^{\mu}(y_{s})=A^{\mu}(y_{s})
\end{equation}
and
\begin{equation}
\not\!D_{s}\psi_{s}(y_{n})=0,\quad \psi_{s}(y_{s})=\psi(y_{s})
\end{equation}
where
\begin{equation}
D_{s\mu}=\partial^{\mu}-ig A_{s}^{\mu},\quad G_{s}^{\mu\nu}=\frac{i}{g}[D_{s}^{\mu},D_{s}^{\nu}]
\end{equation}
The equation should be solved perturbatively, that is, we define soft fields according to perturbation theory. If we choose a suitable gauge in the region $y_{s}$, then momenta of the classical fields in the region $y_{s}$ are all soft. In such gauge, momenta of the classical fields $\psi_{s}$ and $A_{s}$ in the region $y_{n}$ are also soft at lowest order in the coupling constant $g$ as momenta of propagators do not change in the prorogation. At higher order, we assume that discrepancy between rapidity of collinear and soft particles are so large that couplings between soft particles do not produce collinear particles. Thus momenta of the classical fields $\psi_{s}$ and $A_{s}$ in the region $y_{n}$ are soft in higher order in $g$.
We then define the effective fields:
\begin{equation}
\psi_{n}(y_{n})=\psi(y_{n})
\end{equation}
\begin{equation}
\psi_{n}(y_{m})=\psi_{n}(y_{s})=0 (m^{\mu}\neq n^{\mu})
\end{equation}
\begin{equation}
A_{n}^{\mu}(y_{n})=A^{\mu}(y_{n})-A_{s}^{\mu}(y_{n})\end{equation}
\begin{equation}
 A_{n}^{\mu}(y_{m})=A_{n}^{\mu}(y_{m})=0 (m^{\mu}\neq n^{\mu})
\end{equation}

We can then write the Lagrangian density of QCD as:
\begin{eqnarray}
\mathcal{L}(y)&=&\sum_{n^{\mu}}\mathcal{L}_{n}(y)
+\mathcal{L}_{s}(y)
\nonumber\\
&=&\sum_{n^{\mu}}i\bar{\psi}_{n}(\not\!\partial-ig\not\!A_{n}-ig\not\!A_{s})\psi_{n}
\nonumber\\
&&
-\frac{1}{2g^{2}}\sum_{n}^{\mu}tr_{c}\{[\partial^{\mu}-igA_{n}^{\mu}-igA_{s}^{\mu},
\nonumber\\
&&
\partial^{\nu}-igA_{n}^{\nu}-igA_{s}^{\nu}]^{2}
\}
\nonumber\\
&&+i\bar{\psi}_{s}(\not\!\partial-ig\not\!A_{s})\psi_{s}
\nonumber\\
&&
-\frac{1}{2g^{2}}tr_{c}\{[\partial^{\mu}-igA_{s}^{\mu},
\partial^{\nu}-igA_{s}^{\nu}]^{2}\}
\end{eqnarray}
Momenta of the effective fields $\psi_{n}$ and $A_{n}$ can be soft. However, couplings between soft fields constrained in the region $y_{n}$ are power suppressed. To see this, we notice that $\Delta (n\cdot y_{n})\sim 1/Q$ for the region $y_{n}$, while for the coupling between soft particles we have $\Delta y_{s}^{\mu}\sim 1/M$. Thus couplings between soft gluons and soft fermions constrained in the region $y_{n}$ are power suppressed. Couplings between soft fermions and collinear fermions are also power suppressed and can be dropped. While dealing with the coupling between $A_{s}$ and $\psi_{n}$($A_{n}$), we can treat the momenta of $\psi_{n}$ and $A_{n}$ as collinear to $n^{\mu}$. We can then take the eikonal line approximation in such coupling, we have:
\begin{equation}
\partial^{\mu}-igA_{n}^{\mu}-igA_{s}^{\mu}
\simeq \widetilde{\partial}_{n}^{\phantom{n}\mu}-ig A_{n}^{\mu}-ign\cdot A_{s}\bar{n}^{\mu}
\end{equation}
where
\begin{equation}
\widetilde{\partial}_{n}^{\phantom{n}\mu}\psi_{n}=\partial^{\mu}\psi_{n}, \quad
\widetilde{\partial}_{n}^{\phantom{n}\mu}A_{n}^{\nu}=\partial^{\mu}A_{n}^{\nu}
\end{equation}
\begin{equation}
\widetilde{\partial}_{n}^{\phantom{n}\mu}n\cdot A_{s}=\bar{n}^{\mu}n\cdot\partial n\cdot A_{s}
\end{equation}
We then have:
\begin{equation}
\widetilde{\partial}_{n}^{\phantom{n}\mu}-ig A_{n}^{\mu}-ign\cdot A_{s}\bar{n}^{\mu}
=Y_{n}\widetilde{\partial}_{n}^{\phantom{n}\mu}Y_{n}^{\dag}-ig A_{n}^{\mu}
\end{equation}
where
\begin{equation}
\label{soft Wilson line}
Y_{n}(x_{n})=P\exp(ig\int_{-\infty}^{0}\ud s n\cdot A_{s}(x_{n}+sn)
\end{equation}
The Wilson line travel from $-\infty$ to $x_{n}$, this is in accordance with our deformation of integral path of soft gluons. We redefine the fields:
\begin{equation}
\psi_{n}^{(0)}=Y_{n}^{\dag}\psi_{n}
,\quad A_{n}^{(0)\mu}=Y_{n}^{\dag}A_{n}^{\mu}Y_{n}
\end{equation}
and write $\mathcal{L}_{n}$ as:
\begin{eqnarray}
\mathcal{L}_{n}^{(0)}&=&i\bar{\psi}_{n}^{(0)}(\widetilde{\not\!\partial}_{n}
-ig\not\!A_{n}^{(0)})\psi_{n}^{(0)}
\nonumber\\
      &&+\frac{1}{2g^{2}}
      tr\left\{([\widetilde{\partial}_{n}^{\phantom{n}\mu}-igA_{n}^{(0)\mu},\right.
\nonumber\\
&&
       \left. \widetilde{\partial}_{n}^{\phantom{n}\nu}-igA_{n}^{(0)\nu}])^{2}\right\}
\end{eqnarray}
We bring in the notation:
\begin{equation}
\mathcal{L}_{\Lambda}=\sum_{n^{\mu}}\mathcal{L}_{n}^{(0)}+\mathcal{L}_{s}
\end{equation}

$\mathcal{L}_{\Lambda}$ is invariant under the gauge transformation of $U_{s}(y)$ :
\begin{equation}
\psi_{s}(y)\to U_{s}\psi_{s}(y),\quad A_{s}^{\mu}(y)\to U_{s}(A_{s}^{\mu}+\frac{i}{g}\partial^{\mu})U_{s}^{\dag}(y)
\end{equation}
\begin{equation}
\psi_{n}^{(0)}(y)\to \psi_{n}^{(0)}(y),\quad
A_{n}^{(0)\mu}(y)\to A_{n}^{(0)\mu}(y)
\end{equation}
where we have constrained that $U_{s}(\infty)=1$. This is just the usual gauge invariance of QCD. To see this, one may consider the transformation of $\psi_{n}$ and $A_{n}$:
 \begin{equation}
\psi_{n}\to U_{s}\psi_{n}^{(0)},\quad
A_{n}^{\mu}\to U_{s}A_{n}^{\mu}U_{s}^{\dag}
\end{equation}
We then consider the fields $\psi=\sum_{n\mu}\psi_{n}+\psi_{s}$ and $A=\sum_{n^{\mu}}A_{n}+A_{s}$ and have:
\begin{equation}
\psi\to U_{s}\psi,\quad A^{\mu}\to U_{s}(A^{\mu}+\frac{i}{g}\partial^{\mu})U_{s}^{\dag}
\end{equation}
this is just the usual gauge transformation of QCD.
$\mathcal{L}_{\Lambda}$ is also invariant under the transformation:
\begin{equation}
\psi_{s}\to\psi_{s},\quad A_{s}^{\mu}\to A_{s}^{\mu}
\end{equation}
\begin{equation}
\psi_{n}^{(0)}\to U_{c}\psi_{n}^{(0)}
,\quad
A_{n}^{(0)\mu}\to U_{c}(A_{n}^{(0)\mu}+\frac{i}{g}\widetilde{\partial}_{n}^{\phantom{n}\mu})U_{c}^{\dag}
\end{equation}
This is also connected with the gauge invariance of QCD. To see this we consider the special configuration $\psi_{s}=A_{s}=0$, which is invariant under the transformation $U_{c}$ and have:
\begin{equation}
\psi\to U_{c}\psi,\quad A^{\mu}\to U_{c}(A^{\mu}+\frac{i}{g}\partial^{\mu})U_{c}^{\dag}
\end{equation}

We now consider the quantization of $\mathcal{L}_{\Lambda}$. We first inspect the boundary condition given by the region $C(x)$. In $C(x)$, there are not interactions between different jets and soft hadrons as they are color singlets and separate from each other. We denote the region which jet collinear to $n^{\mu}$ and soft hadrons locate in as $x_{n}$ and $x_{s}$.  We have $\psi_{s}(x_{n})=A_{s}(x_{n})=0$ in this case. That is to say $\psi(x_{n})=\psi_{n}^{(0)}(x_{n})$ and $A_{n}=A_{n}^{(0)}(x_{n})$. Thus boundary conditions provided by the region $C(x)$ are constraints on the configuration of the classical fields $\psi_{n}^{(0)}$, $A_{n}^{(0)}$, $\psi_{s}$ and $A_{s}$. We can thus quantize such fields instead of the parton fields $\psi$ and $A$. While dealing with process between soft particles and particles collinear to $n^{\mu}$, such quantization scheme gives the same result as that in QCD at leading order in $M/Q$.  At tree level, this is obvious according to the construction of the classical lagrangian density $\mathcal{L}_{\Lambda}$. At one loop level, the loop momenta are hard, collinear to $n^{\mu}$ or soft on LPSS. If the loop momenta are soft, then one can simply take the eikonal line approximation in the coupling between the internal lines and the collinear external line. Such part can be described by tree level of $\mathcal{L}_{n}^{(0)}$ and one loop level of $\mathcal{L}_{s}$. If the loop momenta are hard or collinear to $n^{\mu}$, then one can drop the momenta components $\bar{n}\cdot q$ and $\vec{q}_{n\perp}$ in these internal lines, where $q$ denote the momenta of soft external lines. That is, one can take the eikonal line approximation in the coupling between the soft external lines and the internal lines. Such part can be described by tree level of $\mathcal{L}_{s}$ and one loop level of $\mathcal{L}_{n}^{(0)}$. One can simply extend this procedure to higher orders in $g$.

There are hard process in the space time region $H(x,1/Q)$, thus $\mathcal{L}_{\Lambda}$ is not enough to describe physics in $H(x,1/Q)$.
LPSS with disconnected hard vertex cancel out after the summation over gauge invariant set of graphs is performed \cite{LS1985}. Thus hard process between different jets can be described by a effective hard vertex. We bring in the effective operator $\Gamma^{\mu}(x)B_{\mu}(x)$ to describe such effects, where $B^{\mu}$ denote the photon field. $\Gamma^{\mu}$ is determined by requiring that it can give the the equivalent hadronic tensor as (\ref{hadronic tensor}) at leading order in $M/Q$. According to the gauge invariance under $U_{s}$, we have:
\begin{equation}
\Gamma^{\mu}(x)=\Gamma^{\mu}(Y_{n}\psi_{n}^{(0)}(x_{n}),\ldots,Y_{m}A_{m}^{(0)\mu}
Y_{m}^{\dag}(x_{m}))(x)
\end{equation}
where $x_{n}^{0}\leq x^{0}$, $x_{m}^{0}\leq x^{0}$ as there are not interactions after the production of hard photon in (\ref{hadronic tensor}).  Couplings between soft particles and modes with $|k^{2}|\gtrsim Q^{2}$ are suppressed as power of $M/Q$, thus we can drop the $\psi(x_{s})$ and $A(x_{s})$ terms in $\Gamma^{\mu}$.

It is convenient to work with the boundary condition $A^{\mu}(\infty)=0$.  We then expand $\Gamma^{\mu}$ according to the small momenta components $n\cdot p_{n}$ and $(p_{n})_{n\perp}$:
\begin{eqnarray}
\Gamma^{\mu}(x)
&=&\sum_{n^{\mu},\ldots, m^{\mu}}\int\ud (n\cdot x_{n})\int\ud (m\cdot x_{m})
\nonumber\\
&&
\mathcal{J}^{\mu}(\psi_{n}(x+n\cdot x_{n}\bar{n}),
\nonumber\\
&&
\ldots,
A_{m}^{\mu}
(x+m\cdot x_{m}\bar{m}))(x)
\nonumber\\
&&
+O(M/Q)
\end{eqnarray}
Contributions that suppressed as powers of $(p_{n})_{n\perp}/\bar{n}\cdot p_{n}$ and $n\cdot p_{n}/\bar{n}\cdot p_{n}$ are absorbed into matching of the effective operators at higher order of $\alpha_{s}$ with corrections of order $M/Q$. The condition $(x+x_{n})^{0}\leq x^{0}$  is equivalent to $n\cdot x_{n}\leq 0$ in this effective action as $\bar{n}\cdot x_{n}=0$.

Scalar polarized collinear gluons should decouple from physical processes according to gauge invariance. We define the fields:
\begin{equation}
\widetilde{A}_{n,x}^{(0)\mu}(x_{n})
=A_{n}^{(0)\mu}(\bar{n}\cdot x,n\cdot x_{n},(\vec{x}_{0})_{n\perp})
\end{equation}
\begin{equation}
\widetilde{A}_{n,x}^{\mu}(x_{n})
=A_{n}^{\mu}(\bar{n}\cdot x,n\cdot x_{n},(\vec{x}_{0})_{n\perp})
\end{equation}
We also bring in the Wilson lines:
\begin{equation}
W_{n,x}^{(0)}(x_{n})=P\exp(ig\int_{-\infty}^{0}\ud s \bar{n}\cdot \widetilde{A}_{n,x}^{(0)}(x_{n}+s\bar{n}))
\end{equation}
\begin{equation}
W_{n,x}(x_{n})=P\exp(ig\int_{-\infty}^{0}\ud s
\bar{n}\cdot \widetilde{A}_{n,x}(x_{n}+s\bar{n})
\end{equation}
They travel from $-\infty$ to $x$ as there are not interactions after the hard collision in the hadronic tensor (\ref{hadronic tensor}). The fields $A^{\mu}(x^{0}+\infty,\vec{y})$ do not contribute to the hadronic tensor (\ref{hadronic tensor}). Similar Wilson lines can be found in literatures, e.g.  \cite{B1985,CSS1985,CSS1988,SCET,CS1981B,QS1991}.
For future convenience, we define the fields:
\begin{eqnarray}
\widehat{\psi}_{n,x}^{(0)}(x_{n})&=&W_{n,x}^{(0)\dag}(x_{n})\psi_{n}^{(0)}(x_{n})
\nonumber\\
(\partial^{\mu}-ig\widehat{A}_{n,x}^{(0)\mu})&=&
W_{n,x}^{(0)\dag}
(\partial^{\mu}-igA_{n}^{(0)\mu})
W_{n,x}^{(0)}
\end{eqnarray}
\begin{eqnarray}
\widehat{\psi}_{n,x}(x_{n})&=&W_{n,x}^{\dag}(x_{n})\psi_{n}(x_{n})
\nonumber\\
(\partial^{\mu}-ig\widehat{A}_{n}^{\mu})(x_{n})&=&
W_{n,x}^{\dag}
(\partial^{\mu}-igA_{n,x}^{\mu})
W_{n,x}
\end{eqnarray}

We consider a special part of $\Gamma^{\mu}$, in which all fields are physical fields. We denote such part as $\Gamma_{phy}^{\mu}$. According to gauge invariance, we have:
\begin{equation}
\Gamma_{phy}^{\mu}=\Gamma_{phy}^{\mu}(\psi_{n},\ldots \partial^{m\perp}-ig A_{m}^{m\perp})
\end{equation}
Contributions of the fields $n\cdot A_{n}$ to $\Gamma^{\mu}$ are power suppressed and can be dropped  according to the power counting displayed in table\ref{Power couting}. We can expand $\Gamma^{\mu}$ according to the fields $\bar{l}\cdot A_{l}$ for all directions $l^{\mu}$. Then the lowest order result equal to $\Gamma_{phy}^{\mu}$. According to such expansion, if we take the lowest perturbation of $\bar{n}\cdot A_{n}$ for a special direction $n^{\mu}$, we have:
\begin{equation}
\Gamma^{\mu}|_{\bar{n}\cdot A_{n}=0}=\Gamma^{\mu}(\psi_{n},\partial^{n\perp}-ig A_{n}^{n\perp}\ldots ,A_{m})
\end{equation}
where $\bar{n}\cdot A_{n}=0$ should be treated as the lowest perturbation of $\bar{n}\cdot A_{n}$ not  the axial gauge.
We then choose a transformation $U_{c}^{\prime}$ so that $U_{c}^{\prime}(\infty)=1$
and $\bar{n}\cdot A(x)=0$ at finite point in this gauge. In this special gauge, we have:
\begin{eqnarray}
\Gamma^{\mu}
&=&\Gamma^{\mu}(\widehat{\psi}_{n,x},\ldots,
A_{m})
\end{eqnarray}
$\widehat{\psi}_{n,x}(x+n\cdot x_{n}\bar{n})$ and
$(\partial^{n\perp}-ig\widehat{A}_{n,x}^{n\perp})(x+n\cdot x_{n}\bar{n})$ are invariant under $(U_{c}^{\prime})^{-1}$.
According to the gauge symmetry of $\Gamma^{\mu}$, we have:
\begin{eqnarray}
\Gamma^{\mu}
&=&\Gamma^{\mu}(\widehat{\psi}_{n,x},\ldots,
A_{m})
\end{eqnarray}
in any gauge with the boundary condition $A(\infty)=0$. We repeat the arguments and have:
\begin{eqnarray}
&&\Gamma^{\mu}|_{A(\infty)=0}
\nonumber\\
&=&\sum_{n^{\mu},\ldots, m^{\mu}}\ldots \int\ud (n\cdot x_{n})\int\ud (m\cdot x_{m})
\nonumber\\
&&
\mathcal{J}^{\mu}(\widehat{\psi}_{n,x}(x+n\cdot x_{n}\bar{n}),
\nonumber\\
&&\ldots,
(\partial^{m\perp}-\widehat{A}_{m,x}^{m\perp})(x+m\cdot x_{m}\bar{m}))
\end{eqnarray}

To make physical fields in $J^{\mu}$ local in $x$, we redefine the collinear fields and extract large momenta components of these fields. That is:
\begin{equation}
\psi_{n}(y)=\sum_{\bar{n}\cdot p}\psi_{n,\bar{n}\cdot p}(y)e^{-i\bar{n}\cdot p n\cdot y}
\end{equation}
\begin{equation}
A_{n}^{\mu}(y)=\sum_{\bar{n}\cdot p}A_{n,\bar{n}\cdot p}^{\mu}(x)e^{-i\bar{n}\cdot p n\cdot y}
\end{equation}
Then the large momenta components become labels on the effective fields. This is similar with the definition of collinear fields as in \cite{SCET}, but we only extract large momenta components. After this extraction, we can take that $\psi_{n,\bar{n}\cdot p}(x+n\cdot x_{n}\bar{n})\simeq \psi_{n,\bar{n}\cdot p}(x)$ and $A_{n,\bar{n}\cdot p}^{\mu}(x+n\cdot x_{n}\bar{n})\simeq A_{n,\bar{n}\cdot p}(x)$ in $\Gamma^{\mu}$ as $n\cdot x_{n}\sim 1/Q$.  $\Gamma^{\mu}$ can then be written as:
\begin{eqnarray}
\Gamma^{\mu}|_{A(\infty)=0}
&=&
\sum_{n,\bar{n}\cdot p,\ldots,m,\bar{m}\cdot p^{\prime}}
\mathcal{J}^{\mu}((\widehat{\psi}_{n,x})_{\bar{n}\cdot p},\ldots,
\nonumber\\
&&
(\partial^{m\perp}-ig\widehat{A}_{n,x}^{m\perp})_{\bar{m}\cdot p^{\prime}}
)(x)
\nonumber\\
&=&
\sum_{n,\bar{n}\cdot p,\ldots,m,\bar{m}\cdot p^{\prime}}
\mathcal{J}^{\mu}(Y_{n}(\widehat{\psi}_{n,x}^{(0)})_{\bar{n}\cdot p},\ldots,
\nonumber\\
&&
Y_{m}(\partial^{m\perp}-ig\widehat{A}_{n,x}^{(0)m\perp})_{\bar{m}\cdot p^{\prime}}
Y_{m}^{\dag})(x)
\end{eqnarray}
If we consider the configuration  $\bar{n}\cdot \widetilde{A}_{n,x}=0$ (one should notice that $\bar{n}\cdot \widetilde{A}_{n,x}=0$ is not the axial gauge, it is just the lowest perturbation of $\bar{n}\cdot \widetilde{A}_{n,x}$), then $\mathcal{L}_{Q}$ can be matched perturbatively in the on-shell scheme.

Contributions of  hard diagrams with several physical external lines nearly collinear to the same direction to $n^{\mu}$ are suppressed as powers of $\theta$ according to the power counting displayed in Table\ref{Power couting}, where $\theta$ denotes the angle between the momenta directions of  these external lines. If these external lines connect to the same jet then one can contract the internal lines with momenta square $|k^{2}|\sim Q^{2}\theta^{2}$ in to hard subdiagram with corrections suppressed as powers of $M/(Q\theta)$ and are of order $\theta\ast M/(Q\theta)=M/Q$. We repeat this procedure(this is just the usual matching procedure) and get the conclusion that physical collinear fields in  effective operators in $\Gamma^{\mu}$ should connect to different jets at leading order with corrections of order $M/Q$.    This is in accordance with the result in \cite{S1978}.
\\

%%%%%%%%%%%%%%%%%%%%%%%%%%%%%%%%%%%%%%%%%%%%%%%%%%%%%%%%%%%%%%%%%

\section{Factorization}
\label{factorize}

In this section we finish the proof of QCD factorization of Drell-Yan process.

We write the hadronic tensor in the effective theory:
\begin{eqnarray}
H^{\mu\nu}(q)&=&\sum_{X}\int\ud^{4}x e^{iq\cdot x}
\big<p_{1}p_{2}|\bar{T}\Gamma^{\dag\mu}(x)
\nonumber\\
&&
T\Gamma^{\nu}(0)|p_{1}p_{2}\big>
\end{eqnarray}
where we have made the summation over all possible points at which the hard photon is produced.
We write $\Gamma^{\mu}$ as:
\begin{eqnarray}
\Gamma^{\mu}(x)&=&\sum_{\Gamma_{c}^{\mu},\Gamma_{s}}^{color\quad indices}
\Gamma_{c}^{\mu}(\widehat{\psi}_{n_{i},x}^{(0)},\widehat{\psi}_{m,x }^{(0)},
\nonumber\\
&&
\partial^{l\perp}-ig\widehat{A}_{l,x }^{(0)l\perp})
\Gamma_{s}(Y_{n_{i}},Y_{m},Y_{l})(x)
\end{eqnarray}
where $n_{i}$($i=1,2$) denote the direction of initial collinear particles, $m$ and $l$ denote the direction of final  collinear particles.
We have omitted possible color indices for simplicity.
$\Gamma_{c}^{\mu}$ is multi-linear with $\widehat{\psi}_{n,x}^{(0)}$, $\widehat{\psi}_{m,x}^{(0)}$ and $(\widehat{A}_{l,x })_{l\perp}^{(0)\mu}$ as physical collinear fields in $\Gamma^{\mu}$ connect to different jets at leading order. Wilson lines that appear in $\Gamma_{s}$ depend on type of partons that appear in $\Gamma_{c}$. We notice that $(s_{1}n^{\mu}-s_{2}m^{\mu})^{2}<0$ for $s_{1}<0$, $s_{2}<0$ and $n^{\mu}\neq m^{\mu}$.  Thus, the order of Wilson lines in $\Gamma_{s}$ do not affect the result.  (There are color indices of Wilson lines in $\Gamma_{s}$, they are matrix-elements not matrixes at this step) In addition, we can choose that $n^{\mu}$ slightly space-like so that the time ordering and anti-time ordering operators in Wilson lines can be dropped.

We can drop the momenta of soft particles and small momenta components of collinear particles in the $\delta$-function of momenta conversation for the whole process.  Thus we can set that $x^{\mu}=(n\cdot x,0,\vec{0})$ in the fields $\widehat{\psi}_{n,x}^{(0)}$ and $\widehat{A}_{n,x}^{(0)}$. We can also set that $x=0$ in Wilson lines $Y_{n}(x)$.  One should notice that there are not detected hadrons in the process considered in this paper and intermediate energy scale that prevent we make such approximation do not exist. The part of $H^{\mu\nu}$ that depend on soft fields are:
\begin{equation}
H^{s}(0)=\big<0|\Gamma_{s}^{\dag}(0)\Gamma_{s}(0)|0\big>
\end{equation}
Hadrons are all color singlets and we have:
\begin{equation}
\sum_{j}(Y_{n})_{ij}(Y_{n}^{\dag})_{ji^{\prime}}=\delta_{ii^{\prime}}
,\quad
\sum_{j}(Y_{n}^{\dag})_{ji}(Y_{n})_{i^{\prime}j}
=\delta_{ii^{\prime}}
\end{equation}
\begin{equation}
\sum_{a}(Y_{n}t^{a}Y_{n}^{\dag})_{ij}(Y_{n}t^{a}Y_{n}^{\dag})_{kl}=
\sum_{a}t^{a}_{ij}t^{a}_{kl}
\end{equation}
We then have:
\begin{eqnarray}
H^{\mu\nu}(q)&=&\sum_{X}\int\ud^{4}x e^{iq\cdot x}
\big<p_{1}p_{2}|\bar{T}\Gamma^{\dag\mu}(x)
\nonumber\\
&&
T\Gamma^{\nu}(0)|p_{1}p_{2}\big>|_{\psi_{s}=A_{s}=0}
\end{eqnarray}

We define the annihilation operator
\begin{eqnarray}
\label{fermion}
\widehat{a}_{n,x,p}^{s}&=&\frac{\sqrt{2E_{p}}}{2m}\int\ud^{3}\vec{x}_{n}
                             \bar{u}^{s}(p)\widehat{\psi}_{n,x}(\vec{x}_{n})
                             e^{-i\vec{p}\cdot \vec{x}}
\end{eqnarray}
\begin{eqnarray}
\label{antif}
\widehat{a}_{n,x,p}^{s}&=&-\frac{\sqrt{2E_{p}}}{2m}\int\ud^{3}\vec{x}_{n}
                   \bar{\widehat{\psi}}_{n,x}(\vec{x}_{n})
                   v^{s}(p)e^{-i\vec{p}\cdot \vec{x}}
\end{eqnarray}
for collinear fermions and anti-fermions respectively. One can verify that $\widehat{a}_{n,p}^{s}|0\big>=0$. For collinear gluons,  we define that:
\begin{eqnarray}
\label{gaugeb}
\widehat{a}_{n,x,p}^{j}&=&\frac{i}{\bar{n}\cdot p}\sqrt{2E_{p}}
\int\ud^{3}\vec{x}_{n}e^{-i\vec{p}\cdot \vec{x}}
\bar{n}\cdot\partial\epsilon^{j*}(p)\cdot\widehat{A}_{n,x}(\vec{x}_{n})
\nonumber\\
&=&\frac{i}{\bar{n}\cdot p}\sqrt{2E_{p}}
\int\ud^{3}\vec{x}_{n}e^{-i\vec{p}\cdot \vec{x}}
\epsilon_{\mu}^{j*}(p)
\nonumber\\
&&
W_{n,x}^{\dag}G_{n}^{n\mu}
W_{n,x}(\vec{x}_{n})
\end{eqnarray}
with $j$ denotes different polarizations, where
\begin{equation}
G_{n}^{n\mu}=\frac{1}{-ig}
[\bar{n}\cdot(\partial-igA_{n}),
\partial^{\mu}-igA_{n}^{\mu}]
\end{equation}
we have made use of the fact $\bar{n}\cdot \widehat{A}_{n,x}(x)=0$. One can also verify that $\widehat{a}_{n,x,p}^{i}|0\big>=0$. Hadron states can then be expanded according to the parton states $|\widehat{p}_{n}\big>_{x}$ which are created by the creation operator $\widehat{a}_{n,x,p}$. That is:
\begin{equation}
|\widehat{p}_{n}\big>_{x}=\sqrt{2E_{p_{n}}}a_{n,x,p_{n}}^{\dag}|0\big>
\end{equation}
where $\widehat{a}_{n,x,p_{n}}^{\dag}$ denote the conjugation of the operators (\ref{fermion}), (\ref{antif}) or (\ref{gaugeb}).

$\Gamma^{\mu}$ is multi-linear with $\widehat{\psi}_{n,x}$ and $\partial^{\mu}-ig \widehat{A}_{m,x}^{\mu}$ with $n\neq m$. There can be no more than one parton states that are created by the operator $\widehat{a}_{n_{i},x,p}^{\dag}$ contracted with $\Gamma{\mu}$ for each initial hadron $p_{i}$.    We denote these active partons as $|\widehat{p}_{i}^{1}\big>_{x}$ and have:
\begin{eqnarray}
\label{factorization}
&&H^{\mu\nu}(q)
\nonumber\\
&=&\lim_{T\to\infty}\sum_{p_{1}^{1},p_{2}^{1}}
2E_{p_{2}^{1}}2E_{p_{1}^{1}}\frac{1}{N_{c}^{2}}\frac{1}{D(G)}
\sum_{\Gamma}
\int\ud^{4}x e^{-iq\cdot x}
\nonumber\\
&&
  \prod_{i=1}^{2}\quad _{-}\big<p_{i}|e^{iH_{n_{i}}(x^{0}+T)}
  tr_{c}\{\widehat{a}_{n_{i},x,p_{i}^{1}}^{\dag}
\nonumber\\
&&
   e^{-iH_{n_{i}}x^{0}}
  \widehat{a}_{n_{i},0,p_{i}^{1}}\} e^{-iH_{n_{i}}T}|p_{i}\big>_{-}|_{\psi_{s}=A_{s}=0}
   \nonumber\\
&& tr_{c}\quad_{x}\big<\widehat{p}_{1}^{1}\widehat{p}_{2}^{1}|
   \Gamma^{\mu\dag}(\vec{x})e^{-i\sum_{n^{\mu}}^{n\neq
    n_{1},n_{2}}H_{n}x^{0}}
\nonumber\\
&&
  \Gamma^{\nu}(\vec{0})|\widehat{p}_{1}^{1}\widehat{p}_{2}^{1}\big>_{0}|_{\psi_{s}=A_{s}=0}
\end{eqnarray}
where $\frac{1}{D(G)}$ denote the color factors produced by fields that do not collinear to initial hadrons. We notice that the states $|\widehat{p}_{n}\big>_{x}$ is invariant under the gauge transformation $U_{n}(x_{n})$($U_{n}(\infty)=1$):
\begin{equation}
\psi_{n}\to U_{n}\psi_{n},\quad
A_{n}^{\mu}\to U_{n}(A_{n}^{\mu}+\frac{i}{g}\partial^{\mu})U_{n}^{\dag}
\end{equation}
$\Gamma^{\mu}$ is also invariant under such gauge transformation even if we choose different $U_{n}$ for different directions $n^{\mu}$. Especially, we can choose that $U_{n}=1$ for $n\neq n_{i}$($i=1,2$). The matrix-elements between parton states in (\ref{factorization}) are invariant under such gauge transformations.  We can  take the lowest perturbation of the fields $\bar{n}\cdot \widetilde{A}_{n,x}$ in such matrix-element, as $\bar{n}\cdot \widetilde{A}_{n,x}$ are scalar polarized.  We then have:
\begin{eqnarray}
\label{final}
&&H^{\mu\nu}(q)
\nonumber\\
&=&\lim_{T\to\infty}\sum_{p_{1}^{1},p_{2}^{1}}
2E_{p_{2}^{1}}2E_{p_{1}^{1}}\frac{1}{N_{c}^{2}}\frac{1}{D(G)}
\sum_{\Gamma}
\int\ud^{4}x e^{-iq\cdot x}
\nonumber\\
&&
  \prod_{i=1}^{2}\quad _{-}\big<p_{i}|e^{iH_{n_{i}}(x^{0}+T)}
   tr_{c}\{\widehat{a}_{n_{i},x,p_{i}^{1}}^{\dag}
\nonumber\\
&&
   e^{-iH_{n_{i}}x^{0}}
  \widehat{a}_{n_{i},0,p_{i}^{1}}\}e^{-iH_{n_{i}}T}|p_{i}\big>_{-}|_{\psi_{s}=A_{s}=0}
   \nonumber\\
&& tr_{c}\big<p_{1}^{1}p_{2}^{1}|
   \Gamma^{\mu\dag}(\vec{x})e^{-i\sum_{n^{\mu}}^{n\neq
    n_{1},n_{2}}H_{n}x^{0}}
\nonumber\\
&&
  \Gamma^{\nu}(\vec{0})|p_{1}^{1}p_{2}^{1}\big>
|_{\psi_{s}=A_{s}=\bar{n}_{i}\cdot \widetilde{A}_{n_{i},x}=\bar{n}_{i}\cdot \widetilde{A}_{n_{i},0}=0}
\end{eqnarray}
where $|p_{i}^{1}\big>$ is the usual partons produced by the operator $a_{p_{i}}^{\dag}=\widehat{a}_{n_{i},x,p_{i}}^{\dag}|_{\bar{n}\cdot \widetilde{A}_{n_{i},x}=0}$. The condition $\bar{n}\cdot \widetilde{A}_{n,x}=0$ should be treated as the lowest perturbation of the fields $\bar{n}\cdot \widetilde{A}_{n,x}$ not the axial gauge.

The formula (\ref{final}) is our final result. One can easily see that soft gluons and scalar polarized gluons that collinear to initial hadrons decouple from the matrix-element between parton states in (\ref{final}). Such matrix-element can be calculated according to perturbation theory.

%%%%%%%%%%%%%%%%%%%%%%%%%%%%%%%%%%%%%%%%%%%%%%%%%%%%%%%%%%%%%%%%%
\section{Conclusion}
\label{conclusion}

We have finished the proof of factorization theorem for Drell-Yan process at operator level. Be different from conventional diagram-level approach \cite{B1985,CSS1985,CSS1988}, we use unitarity of the time evolution operator of QCD direct to cancel interactions after the collision in the hadronic tensor. Time evolution of electro-magnetic currant is free of LPSS caused by Glauber gluons after such cancellation.  Decoupling of soft gluons from collinear jets is realized by defining collinear fields that decouple from soft gluons.

We give some discussion here:

(1)We treat partons of hadrons in different jets as different particles. This is the consequence of the fact that different jets moving in different directions and separate from each other in coordinate space. The interference effects between different jets are described by the interactions among particles in different jets and interactions between collinear and soft particles.

(2)We have dropped the coordinates $\vec{x}_{n\perp}$ and $\bar{n}\cdot x$ of fields collinear to $n^{\mu}$ in (\ref{final}). Thus the parton distribution part of $(\ref{final})$ and the $\delta$-function of momenta conservation are independent of $\vec{x}_{n_{i}\perp}$ and $\bar{n}_{i}\cdot x$ with corrections of order $M/Q$. There are not internal lines with momenta collinear to $n_{i}^{\mu}$($i=1,2$) in the matrix-element between parton states in (\ref{final}).  Such matrix-element can rely on $(\vec{p}_{i}^{1})_{n_{i}\perp}$ and $n_{i}\cdot p_{i}^{1}$ only through the initial external lines. We can drop these terms in such matrix-element with corrections of order $M/Q$.

(3) To get the similar form as in \cite{B1985,CSS1985,CSS1988}, one should  factorize the spinor structure and consider the renormalization of the parton distribution functions. We do not show this here.

%%%%%%%%%%%%%%%%%%%%%%%%%%%%%%%%%%%%%%%%%%%%%%%%%%%%%%%%%%%%%%%%%
\section*{Acknowledgements}

We would like to acknowledge G. T. Bodwin, Y. Q. Chen and J. Qiu for  helpful discussions and important suggestions about the article. This work was supported by the National Nature Science Foundation of China under grants No.11275242.

%%%%%%%%%%%%%%%%%%%%%%%%%%%%%%%%%%%%%%%%%%%%%%%%%%%%%%%%%%%%%%%%%


\begin{thebibliography}{}

\bibitem{BBL1981} G. T. Bodwin, S. J. Brodsky and G. P. Lepage, Phys. Rev. Lett.47(1981)1799.

\bibitem{M1982} A. H. Mueller, Phys. Lett. 108B(1982)355.

\bibitem{LRS1982} W. W. Lindsay, D. A. Ross and C. T. Sachrajda, Nucl. Phys. B214(1982)388.

\bibitem{CSS1984} J. C. Collins, D. E. Soper and G. Sterman. Phys. Lett. 134B(1984)263.

\bibitem{B1985} G. T. Bodwin, Phys. Rev. D31(1985)2616; Erratum-ibid. D34(1986)3932.

\bibitem{CSS1985} J. C. Collins, D. E. Soper and G. Sterman, Nucl. Phys. B261(1985)104.

\bibitem{CSS1988} J. C. Collins, D. E. Soper and G. Sterman, Nucl. Phys. B308(1988) 833.

\bibitem{CS1981B} J. C. Collins and D. E. Soper, Nucl. Phys. B183(1981)381.

\bibitem{CS1981S} J. C. Collins and G. Sterman, Nucl. Phys. B185(1981)172.

\bibitem{S1978}   G. Sterman, Phys. Rev. D17(1978)2773, ibid. 2789.

\bibitem{SCET} C. W. Bauer, S. Fleming, D. Pirjol and I. W. Stewart, Phys. Rev. D63(2001)114020;
               C. W. Bauer, D. Pirjol and I. W. Stewart, Phys. Rev. D65(2002)054022.

\bibitem{QS1991} J. Qiu and G. Sterman,  Nucl. Phys. B353(1991)105.

\bibitem{LS1985} J. M. F. Labastida and G. Sterman, Nucl. Phys. B254(1985)425.


\end{thebibliography}
\end{document}